\documentclass[a4paper,11pt]{article}
\pdfoutput=1 % if your are submitting a pdflatex (i.e. if you have
\usepackage{jinstpub} % for details on the use of the package, please
\usepackage{lineno}
\usepackage{booktabs} %formato da tabela
\usepackage{caption} %Legenda
\usepackage{graphicx}
\usepackage{multicol}
\title{\boldmath Machine Learning for Real-Time Processing of ATLAS Liquid Argon
Calorimeter Signals with FPGAs\footnote{Copyright [2018] CERN for the benefit of the [ATLAS Collaboration]. CC-BY-4.0 license.}}

\author[a]{N. CHIEDDE}

\affiliation[a]{Aix-Marseille Université, CNRS/IN2P3, CPPM,\\Av. de Luminy 163, 13009 Marseille - France}  

\emailAdd{chiedde@cppm.in2p3.fr}

\abstract{ 

The ATLAS experiment at CERN measures energy of proton-proton (p-p) collisions  with a repetition frequency of 40 MHz at the Large Hadron Collider (LHC). The readout electronics of liquid-argon (LAr) calorimeters are being prepared for high luminosity-LHC (HL-LHC) operation as part of the phase-II upgrade, anticipating a pileup of up to 200 simultaneous p–p interactions. The increase of the number of p-p interactions implies that calorimeter signals of up to 25 consecutive collisions overlap, making energy reconstruction more challenging.  

In order to achieve the goal of the HL-HLC, field-programmable gate arrays (FPGAs) are used to process digitized pulses sampled at 40 MHz in real time and different machine learning approaches are being investigated to deal with signal pileup. The convolutional and recurrent neural networks outperform the optimal signal filter currently in use, both in terms of assigning the reconstructed energy to the correct proton bunch crossing and in terms of energy resolution. The enhancements are focused on energy obtained from overlapping pulses. Because the neural networks are implemented on an FPGA, the number of parameters, resource usage, latency and operation frequency must be carefully analysed. A very good agreement is observed between neural network implementations in FPGA and software.  

}


\collaboration[c]{on behalf of the ATLAS Liquid Argon calorimeter group}

\proceeding{N$^{\text{th}}$ Workshop on Electronics for Particles Physics\\
  20 – 24 Sep 2021\\
  On-line}

\begin{document}

\maketitle
\flushbottom

\section{Introduction}
\label{sec:intro}
The ATLAS detector~\cite{a} is placed at the Large Hadron Collider~\cite{b} (LHC) and is used to detect particles generated in high-energy p-p collisions. Every 25 ns, the proton bunches collide, resulting in a collision frequency of 40 MHz. Scheduled to begin with Run-4 in 2027, the next high-luminosity phase of the LHC (HL-LHC) is projected to achieve instantaneous luminosities of 5-7x$10^{34}cm^{-2}s^{-1}$. This corresponds to 140–200 p-p interactions occurring at the same time. The ATLAS liquid argon (LAr) calorimeter mainly exploit the ionisation signal to measure the energy of electromagnetic showers of photons, electrons, and positrons. The fact that up to 25 signal pulses produced in successive LHC bunch crossings (BCs) might overlap, resulting in an out-of-time pileup, significantly decrease the energy resolution of the LAr calorimeter.

Each of the 182,000 calorimeter cells is required to reconstruct the deposited energy at the correct BC with the high energy resolution. The calorimeter is expected to provide real time energy reconstruction to the ATLAS trigger system, thus continuous data processing is required. As a result, the digital processing of LAr calorimeter signals in run 4 must be able to manage continuous data. Due to the huge input data bandwidth of about 250 Tbps delivered through serial connections with 36,000 optical fibers, FPGA technology was chosen over alternative processing devices.  In the current design options, one Intel Stratix-10 FPGA~\cite{c}, with a latency requirement of about 150 ns ~\cite{d,e}, will process 384 or 512 LAr calorimeter cells, which corresponds to data measured by three or four so-called front-end boards (FEBs), respectively.  

\section{Energy reconstruction in the ATLAS liquid argon Calorimeter}
\label{sec:intro}
To calculate the energy in each cell, the current readout electronics of the LAr calorimeters digitize the electronic pulse from the calorimeter at 40 MHz and use an optimal filter~\cite{f} (OF) algorithm. Electronic noise and signal pileup are reduced by using a linear combination of up to five digitized pulse samples. To allocate energy to the right BCs, a peak finder is employed. The optimum filter expects a perfect pulse shape, which leads to reduced performance when the pulse is distorted by prior events. In this case, the peak finder fails to efficiently assign the energy to the correct BC.

We develop Artificial Neural Network (ANN) based methods to improve the energy resolution at the HL-LHC. The ANNs are trained using simulated HL-LHC data obtained by AREUS ~\cite{g}, which includes electronics noise and low-energy deposits in the range up to approximately 1 GeV from particles produced in inelastic  p–p collisions. To  emulate  hard-scattering  events, a uniform transverse energy spectrum is overlaid randomly,  with  maximum  energy  deposits  of  5  GeV at a mean interval of 30 BC with a standard deviation of 10 BC. The simulation is performed for one cell in the barrel section of the LAr calorimeter with an average pileup (\text{µ})=140.

\section{Neural network} 
Two neural networks architectures based on Convolutional and Recurrent Neural Networks (CNNs, RNNs) are evaluated for energy reconstruction. Keras~\cite{m} and TensorFlow~\cite{n} are used to develop and train the ANNs.

\subsection{Convolutional neural networks}

\begin{multicols}{2}

CNNs~\cite{h} are studies as an alternative to the OF approach.
The networks use a sliding-window method to analyse the input data sequence. The best performance is achieved by splitting the CNN architecture into two sub-networks that are tuned for distinct tasks.
The first tagging network structure detects relatively high energy deposits over 3$\sigma$ of electronic noise threshold, which corresponds to 240 MeV.
A detection probability is provided along with the sample sequence to a second structure that is trained to rebuild the deposited energy in each calorimeter cell. 

The two CNNs named 3-Conv and 4-Conv  (figure \ref{CNN}) have the same tagging part configuration, but the energy reconstruction consists of one (3-Conv) or two (4-Conv) convolutional  layers. The OF achieves a maximum signal efficiency of about 80\%, while the tagging CNN reaches efficiencies well above 90\%.
\hspace{5cm}

\begin{center}
\includegraphics[width=0.76\linewidth]{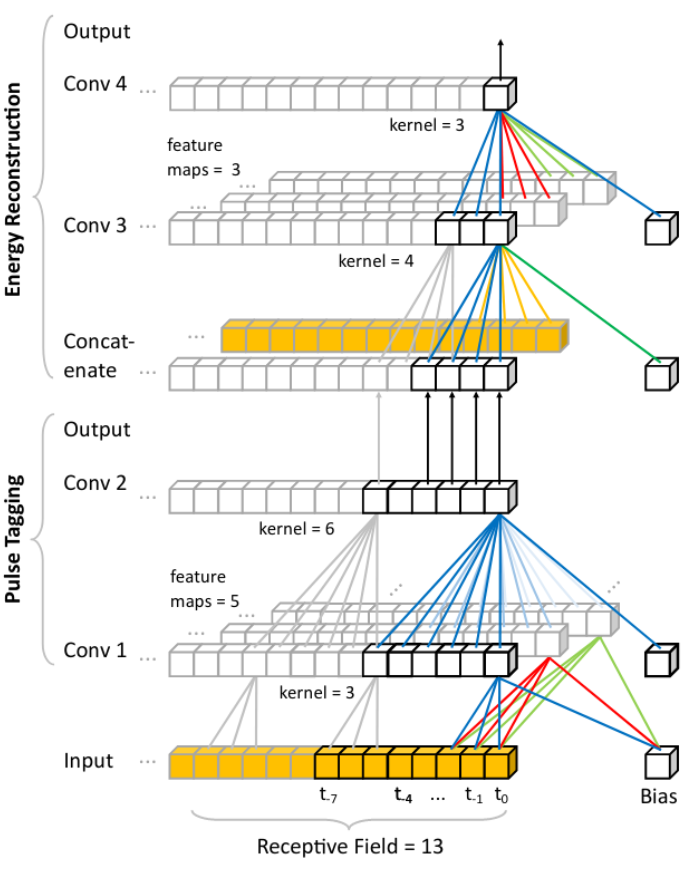}
\captionof{figure}{Representation of the structure of the CNNs developed for energy reconstruction. The pulse tagging part allows to detect energy deposits about the noise level. The energy reconstruction parts uses the pulse tagging output and the digitized input from the calorimeter to reconstruct the energy~\cite{o}.} 
\label{CNN}
\end{center}

\end{multicols}

\subsection{Recurrent neural networks}

\begin{multicols}{2}

RNN are designed to process time series data. It consists of internal neural network that process the input at the current time combined with past processed state. They are excellent candidates for quantifying deposited energy from time-ordered digital LAr signals.
Vanilla-RNN~\cite{i} and long short-term memory (LSTM)~\cite{j}  are the two RNN architectures that are explored. The vanilla-RNN is a network topology with substantially fewer parameters and contains just one activation function, in our case we choose the ReLU activation function. On the other hand, the LSTM has a sophisticated internal structure that uses neural network layers with sigmoid and tanh activation functions to gate the flow of information to the next timestep. As a result, LSTM can process longer sequences and can be used in two different methods. Sliding window  method (figure \ref{slinding_window}), where the digitized signal from the calorimeter is divided into overlapping sub-sequences and each sub-part has a single reconstructed energy. The second method is a single cell, which is a continuous processing of information at each timestamp without the usage of a specified sequence interval. For the Vanilla-RNN, it only works with the sliding window approach due their simplicity.

\begin{center}
\includegraphics[width=0.95\linewidth]{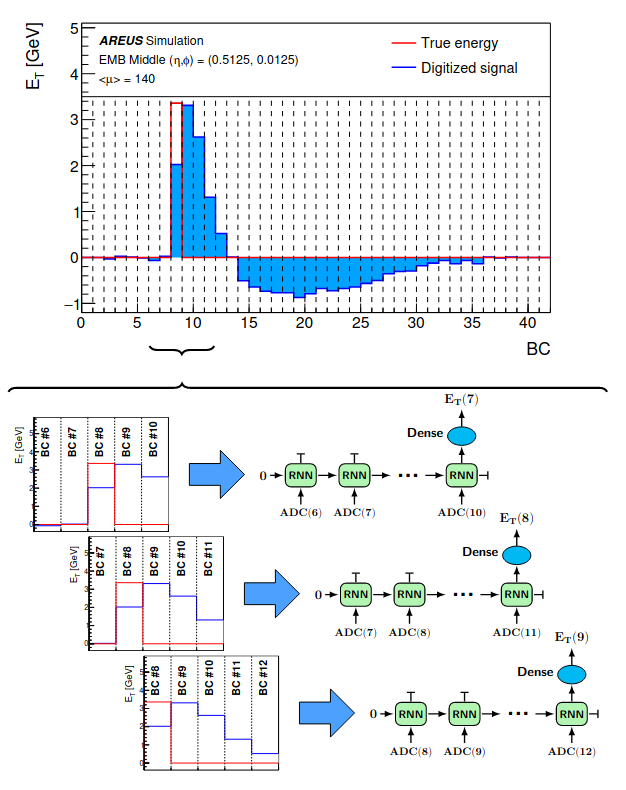}
\captionof{figure}{RNN processing of calorimeter samples with sliding window~\cite{o}.}
\label{slinding_window}
\end{center}

\end{multicols}

\section{Implementation Results} 
\subsection{Neural network performance}

\begin{multicols}{2}
\begin{center}
\includegraphics[width=0.82\linewidth]{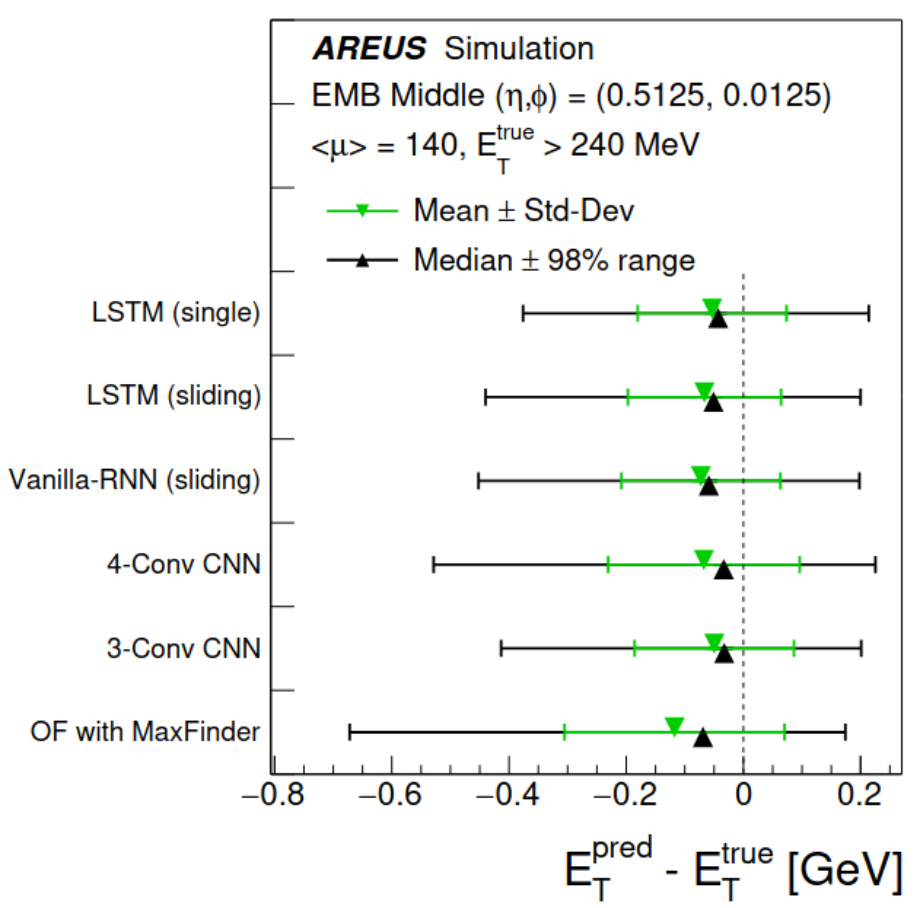}
\captionof{figure}{Energy resolution for different algorithms~\cite{o}.}
\label{simulation}
\end{center}

\begin{center}
\includegraphics[width=0.9\linewidth]{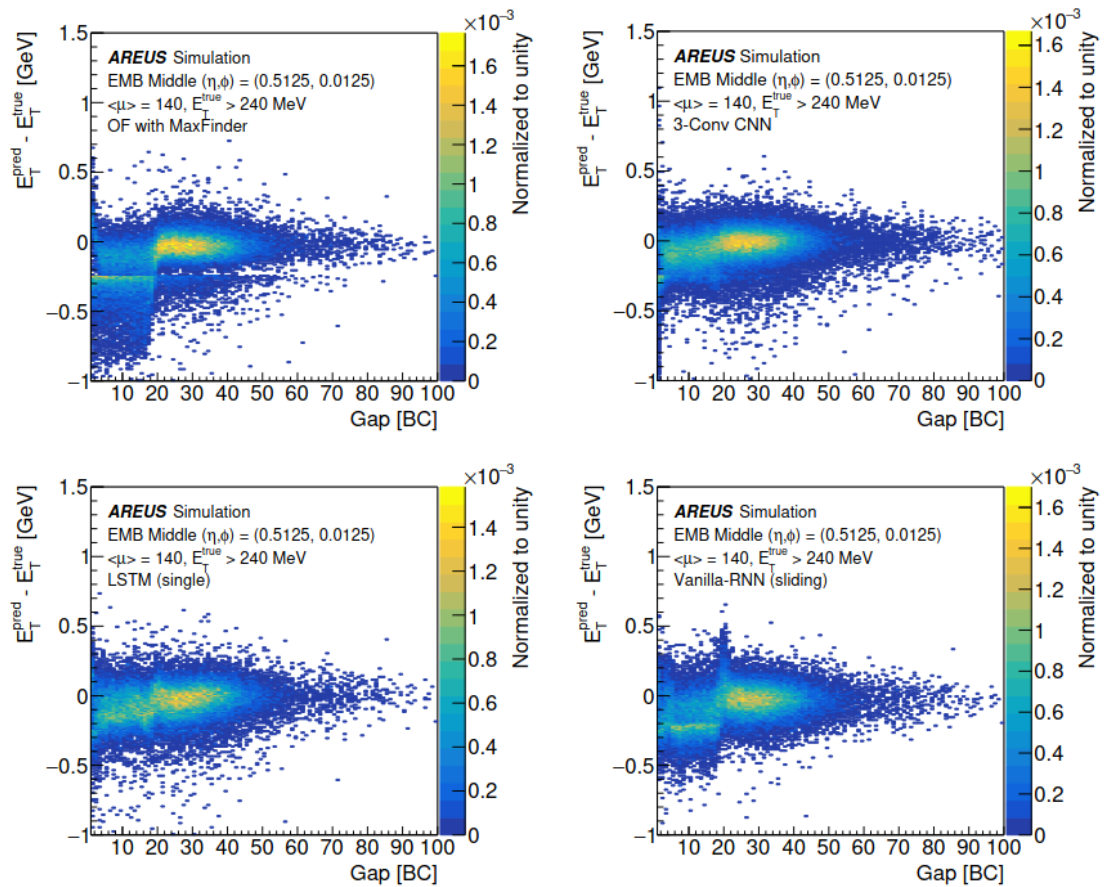}
\captionof{figure}{Resolution as function of the distance to previous high energy deposit for the OF with maximum finder, LSTM, Vanilla-RNN and 3-Conv CNN algorithms~\cite{o}.}
\label{gap}
\end{center}
\end{multicols}

Figure \ref{simulation} shows a comparison of the energy resolution between various NN algorithms and the optimal filtering algorithms. Only energy deposits above 3$\sigma$ the noise thresholds are considered. All four ANNs outperform the OF performance.

Figure \ref{gap} shows the energy resolution as function of the time gap between subsequent energy deposits. At low gap, leading to overlapping pulses, the OF performances degrade significantly. The NNs are capable of recovering the performance in this low gap region.

\subsection{FPGA performence}

\begin{multicols}{2}

The CNNs and RNNs implementations were made with different hardware description languages. Very High-speed integrated circuit hardware Description Language (VHDL) is used to implement CNNs and  High Level Synthesis (HLS) is used for RNNs. Those implementations are simulated in Quartus 20.4 ~\cite{k} and Questa Sim 10.7c ~\cite{l} respectively and their output is compared to the one from Keras. The small differences observed in figure \ref{grafico} are caused by quantization and by the LUT-based realisation of the activation functions. 

\begin{center}
\includegraphics[width=0.7\linewidth]{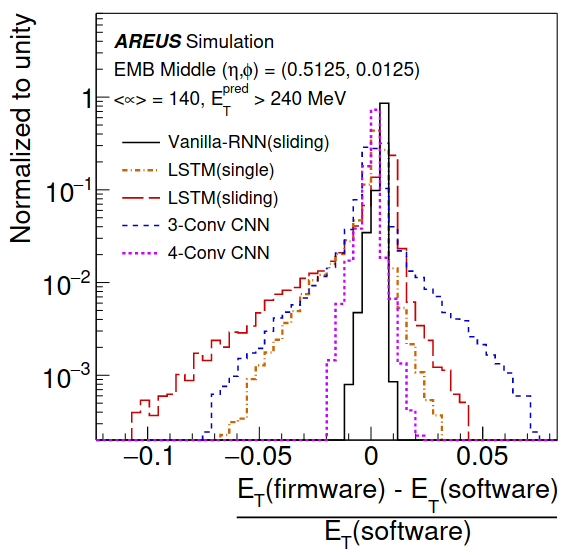}
\captionof{figure}{Relative deviation of the firmware and software results~\cite{o}.}
\label{grafico}
\end{center}

\end{multicols}

 Table \ref{tab:a} shows implementations  on  a  Stratix-10 FPGA for a single data input channel to compare the maximum execution frequency, latency, initiation interval  and  resource  usage  in  terms  of  number  of  digital signal processing (DSP)  and  adaptive logic modules (ALM). The maximum achievable processing frequency for all implementations is in the range of 480–600 MHz.  Receiving at the LHC BC frequency of 40 MHz, it's possible to implement fifteen-fold multiplexing of the input data for vanilla RNN and six-fold for CNNs (Table \ref{tab:i}). The VHDL implementation targets mainly low latency for fast execution. The HLS implementation targets high frequency to allow higher multiplexing. This is reflect in the performance shown in table \ref{tab:i}. Optimization of both implementation is ongoing to find an acceptable compromise between the high frequency and the low latency to fit the readout requirements for the LAr phase-II upgrade.
 
 \hspace{1cm}

\begin{minipage}[b]{0.85\linewidth}
%\centering
\captionof{table}{\label{tab:a} Maximum achievable frequency, latency, initiation interval and resource usage of the VHDL (CNNs) and the HLS (RNNs) implementation on a Stratix-10 FPGA and single channel~\cite{o}.}
\resizebox{12.47cm}{!}{ 
\begin{tabular}{|l|l|l|l|l|c|}
\hline
\textbf{ } & \textbf{\begin{tabular}{c} Max Frequency \\\resizebox{0.9cm}{!}{[MHz]}\end{tabular}} & \textbf{\begin{tabular}{c} Clock \\ Cycle\end{tabular}}  & \textbf{\begin{tabular}{c} Initiation \\ Intreval\end{tabular}} & \textbf{
\begin{tabular}{c} Resource Usage \\ 
\resizebox{1.9cm}{!}{ (DSPs / ALMs)} \end{tabular}}\\
\hline
3-Conv & 493 & 62 & 1 & 0.8\% / 0.6\% \\
4-Conv & 480 & 58 & 1 &  0.7\% / 0.6\% \\
Vanilla (**) & 641 & 206 & 1 & 0.6\% / 1.4\% \\
LSTM (*) & 560 & 220 & 220 & 3.1\% / 1.9\% \\
LSTM (**) & 517 & 363 & 1 & 12.8\% / 7.5\% \\

\hline
\end{tabular}
}
%\end{center}
\\\resizebox{3.3cm}{!}{(*)Single (**)Sliding}
\end{minipage}

\begin{minipage}[b]{1\linewidth}
%\centering
\captionof{table}{\label{tab:i} Occupancy of the NNs implementation on a Stratix-10 FPGA~\cite{o}.}
\smallskip
\resizebox{13cm}{!}{ 
\begin{tabular}{|l|l|l|l|l|c|}
\hline
\textbf{ } & \textbf{Multiplicity} & \textbf{\begin{tabular}{c} Max Frequency \\\resizebox{0.9cm}{!}{[MHz]}\end{tabular}} & \textbf{\begin{tabular}{c} Clock \\ Cycle\end{tabular}} & \textbf{\begin{tabular}{c} Max LAr\\ Channels\end{tabular}} & \textbf{\begin{tabular}{c} Resource Usage \\ 
\resizebox{1.9cm}{!}{ (DSPs / ALMs)} \end{tabular}}\\
\hline
3-Conv & 6 & 344 & 81 & 390 & 0.8\% / 1.5\% \\
4-Conv & 6 & 334 & 62 & 352 &  0.7\% / 1.7\% \\
Vanilla & 15 & 640 & 120 & 576 & 2.6\% / 0.6\% \\
\hline
\end{tabular}
}
\end{minipage}

\acknowledgments

This work was in part supported by the German Federal Ministry of Education and Research within the research infrastructure project 05H19ODCA9. The project leading to this publication has received funding from Excellence Initiative of Aix-Marseille Université - A*MIDEX, a French ”Investissements d’Avenir” programme, AMX-18-INT-006.



\begin{thebibliography}{99}

\bibitem{a}
ATLAS    Collaboration, \emph{The  ATLAS  experiment  at  the  CERN large hadron collider},
JINST 3:S08003,
Publisher (2008).

\bibitem{b}
Evans  L,  Bryant  Ph, \emph{LHC machine},
JINST  3:S08001,
Publisher (2008).

\bibitem{c}
Intel Corporation, \emph{Intel stratix-10 device datasheet},
Version 2020.12.24
Publisher (2020).

\bibitem{d}
ATLAS  Collaboration, \emph{Technical  design  report  for the  phase-II  upgrade  of  the  ATLAS  TDAQ  system, https://cds.cern.ch/record/2285584/},
CERN-LHCC-2017-020, ATLAS-TDR-029,
Publisher (2017).

\bibitem{e}
ATLAS  Collaboration, \emph{Technical  design  report  for  the  phase-II  upgrade  of  the  ATLAS  LAr  calorimeter, https://cds.cern.ch/record/2285582/},
CERN-LHCC-2017-018, ATLAS-TDR-027, 
Publisher (2017).

\bibitem{f}
Cleland WE, Stern EG, \emph{Signal processing considerations for liquid ionization calorimeters in a high rate environment},
NIM A 338:467–49,
Publisher (1994).

\bibitem{g}
Madysa N, \emph{AREUS: a software framework for ATLAS readout electronics upgrade simulation},
EPJ Web Conf 214:02006,
Publisher (2019).

\bibitem{m}
Chollet, Fran\c{c}ois et al., \emph{Keras, https://keras.io},
Accessed (Feb 2021).



\bibitem{n}
Mart\'{\i}n~A, Ashish~A, Paul~B et al., \emph{TensorFlow: Large-Scale Machine Learning on Heterogeneous Systems, https://www.tensorflow.org/},
Accessed (Feb 2021).

\bibitem{h}
LeCun Y et al, \emph{Backpropagation applied to handwritten zip code recognition},
Neural Comput 1(4):541–551,
Publisher (1989).

\bibitem{i}
Sherstinky A, \emph{Fundamentals of recurrent neural network (RNN) and long short-term memory (LSTM) network},
physd. 2019. 132306,
Publisher (2020).

\bibitem{j}
Hochreiter S et al., \emph{Long short-term memory. Neural},
neco. 1997.9. 8. 1735,
Publisher (1997).

\bibitem{k}
Intel, \emph{Quartus, ModelSim and HLS tools,},
https:// www. intel. com., 
Accessed (Feb 2021).

\bibitem{l}
Siemens, \emph{Questa Sim},
Accessed (Jun 2021).

\bibitem{o}
Aad G, Berthold A, Calvet T, Chiedde N et al.,  \emph{Artificial Neural Networks on FPGAs for Real-Time Energy Reconstruction of the ATLAS LAr Calorimeters, https://cds.cern.ch/record/2775033/},
ATL-LARG-PROC-2021-001,
Publisher (2021).


% Please avoid comments such as "For a review'', "For some examples",
% "and references therein" or move them in the text. In general,
% please leave only references in the bibliography and move all
% accessory text in footnotes.

% Also, please have only one work for each \bibitem.


\end{thebibliography}
\end{document}